\documentclass[twocolumn,showpacs,amsmath,amssymb,aps,prb]{revtex4-1}

\usepackage{graphicx}
\usepackage{bm}
\usepackage{verbatim}
\usepackage{amssymb, amsmath}

\usepackage[dvips]{color}

\begin{document}
\title{Energy relaxation in graphene and its measurement with supercurrent}
\author{J.~Voutilainen, A.~Fay, P.~H\"akkinen, J.~K.~Viljas, T.~T.~Heikkil\"a, and P.~J.~Hakonen}
\affiliation{Low Temperature Laboratory, Aalto University, P.O. Box 15100, FI-00076 AALTO, Finland}

\date{\today}

\begin{abstract}
We study inelastic energy relaxation in graphene for low energies to find out how electrons scatter with acoustic phonons and other electrons. By coupling the graphene to superconductors, we create a strong dependence of the measured signal, i.e.,\ critical Josephson current, on the electron population on different energy states. Since the relative population of high- and low-energy states is determined by the inelastic scattering processes, the critical current becomes an effective probe for their strength. We argue that the electron-electron interaction is the dominant relaxation method and, in our model of two-dimensional electron-electron scattering, we find a scattering time $\tau_{e-e}=5\ldots 13$~ps at $T=500$~mK, 1-2 orders of magnitude smaller than predicted by theory.
\end{abstract}

\pacs{72.80.Vp, 74.45.+c, 72.15.Lh}
\maketitle

\section{Introduction}
The special electronic structure of graphene shows up in its electronic properties. \cite{dassarma, eerev} Most attention has been paid to the electronic conductivity \cite{castroneto09} which, due to the strong energy dependence of the density of states, can be tuned significantly with a gate voltage. For dirty graphene, this gate dependence is furthermore modified by elastic scattering \cite{hwang07} due to potential inhomogenities forming charge puddles. At high voltages, also inelastic scattering due to optical phonons appears \cite{meric08}, but the low-energy inelastic scattering due to electron-electron (e-e) \cite{mirlin, mueller} or electron- acoustic phonon scattering (e-ph) \cite{tse, Janne} does not directly influence the conductivity because the mean free paths for them are typically larger than the elastic mean free path. \cite{dassarma, castroneto09}

\begin{figure}[t!]
\centering
\includegraphics[width=0.9\columnwidth]{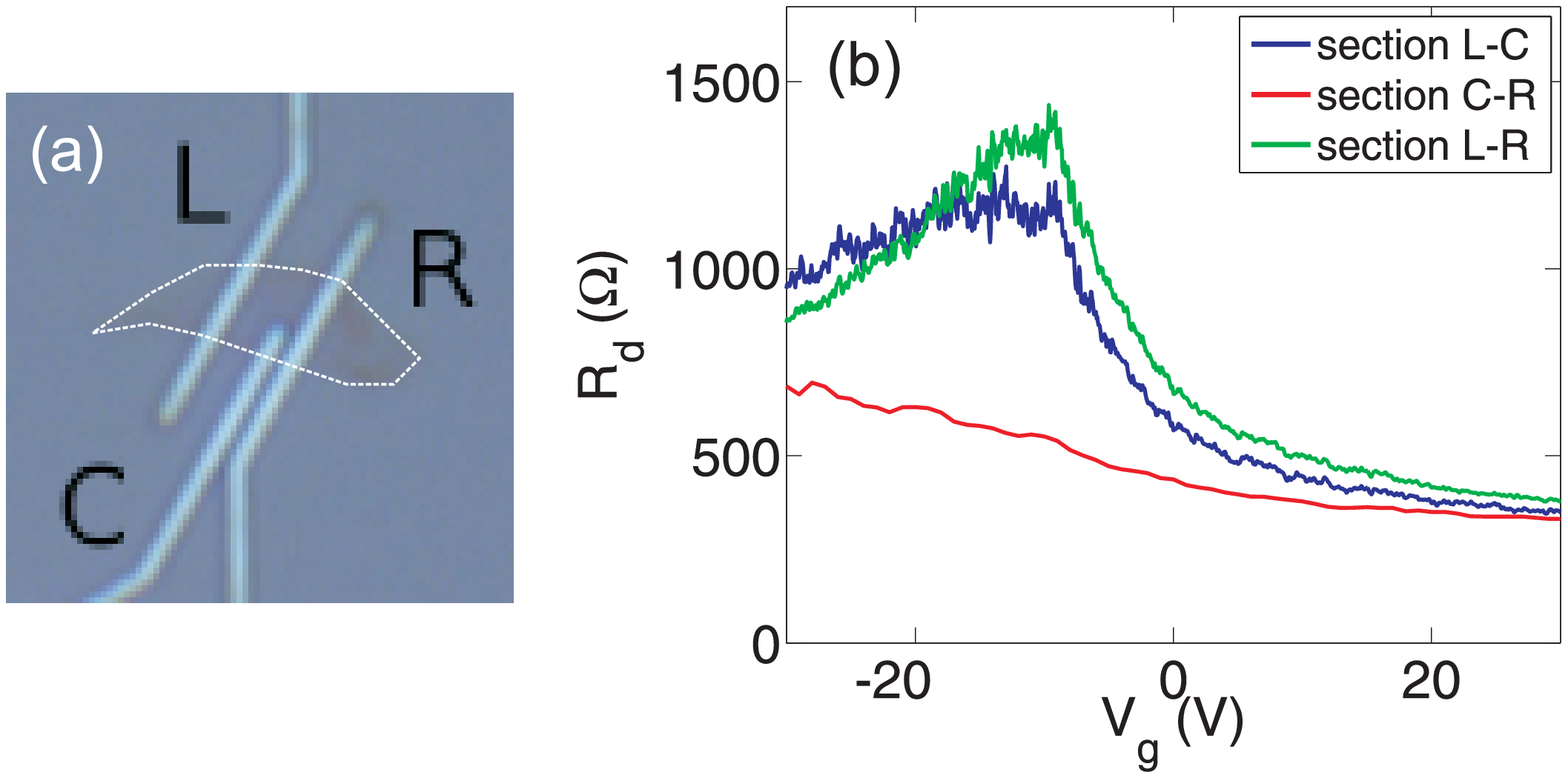}
\caption{(color online). (a) Optical image of the studied graphene flake with three Ti/Al electrodes marked L, C and R. The contour of the flake is highlighted by a a dashed line. (b) Differential resistance of the long (L-C) and short (C-R) sections as a function of gate voltage at $T=4$~K.}
\label{fig:set}
\end{figure}

In this paper we study the effect of low-energy inelastic scattering in graphene by using Josephson critical current measurements to determine heat transport in the system. The idea is to apply a heater voltage and to measure the increased temperature via nearby thermometers. The measurement is performed at sub-kelvin temperatures and low voltages, thereby providing access to low-energy inelastic scattering processes and allowing to disregard scattering from optical phonons. To perform the measurement we use three superconducting electrodes fabricated on graphene. Two of them (thermometer electrodes, C and R in Fig.~\ref{fig:set}) lie close to each other, so that we can measure a finite supercurrent through them, and the third one (heater electrode, L in Fig.~\ref{fig:set}) is used for heating the system. The supercurrent is sensitive to the electron temperature or, more accurately, to the electron distribution function on the graphene region between the superconductors \cite{du08} and therefore acts as an electron thermometer.

We make two measurements: First we measure the supercurrent as a function of temperature, and thus calibrate the thermometer and find parameters for a microscopic theory describing the supercurrent in the junction. Second, we apply a voltage to the heater electrode, supplying heat into the electron system, and measure again the supercurrent in the presence of the heater voltage. The magnitude of the supercurrent in the presence of the heater voltage is sensitive to the strength of inelastic relaxation inside graphene, allowing us to measure it.

The Joule heat generated in the presence of a bias voltage is dissipated either to the electrodes or to phonons.\cite{tse, Janne, kubakaddi09} However, since the electron-acoustic phonon coupling is so weak in graphene, most of the heat escapes into the electrodes, even though this process is blocked at low energies by the superconducting gap $\Delta$.  The escape is possible as a result of processes transferring excitations from low energies to above the gap, in particular multiple Andreev reflections \cite{MAR-S} and e-e scattering. The former tends to broaden the electron distribution below the gap and thus to increase the effective temperature at these energies, while the latter drives the distribution towards a quasiequilibrium (Fermi function) form having a lower effective temperature.  Our thermometer is most sensitive to the distribution at energies below $\Delta$, and therefore it is a sensitive probe of the e-e scattering strength in graphene. But to extract the magnitude of the e-e scattering we have to abandon the simple effective temperature description used for example in Ref.~\onlinecite{roddaroup10} and rather solve a full kinetic Boltzmann equation, with e-e scattering included explicitly with a collision integral.

This paper is organized as follows: In Sec.~\ref{sec:exps}, we describe the experimental setup used to carry out the measurements. In Sec.~\ref{sec:thry}, we formulate our theoretical model and consider the different sources of inelastic relaxation. In Sec.~\ref{sec:res} we combine the theoretical and experimental results to provide an estimate for the strength of relaxation in the system. Finally, we discuss the implications of these results in Sec.~\ref{sec:dis}.

\section{Experiments}\label{sec:exps}
An optical image of the studied monolayer graphene sample is shown in Fig.~\ref{fig:set}(a). The $2.8$ $\mu$m long and $4.0$ $\mu$m wide graphene area in between leads L and R  is partially interrupted by a $1.0$ $\mu$m wide lead C. This latter lead is $1.7$ $\mu$m and $0.4$~$\mu$m far from leads L and R, respectively, with all the distances measured between leads' internal edges. The graphene flake has been exfoliated with a semiconductor wafer dicing tape and deposited on top of a 250~nm thick SiO$_2$ layer. The oxide isolates the graphene flake from a highly {\it p}-doped Si substrate used as a back gate in our experiments. Three Ti/Al (10~nm/50~nm) metallic contacts were patterned by using standard electron-beam lithography techniques, and evaporated in ultra-high vacuum. A $10^{-10}$~mbar vacuum during the metal evaporation guarantees highly transparent contacts, which are needed to observe proximity-induced supercurrents. The sample was measured at low temperatures, down to 80~mK, in a dry dilution cryostat BF-SD250 from Bluefors. The sample contacts are electrically connected to room temperature electronics via one-meter long thermocoaxes, low-pass RC filters (cut-off frequency of 1~kHz) and one-meter twisted pairs, protecting the sample from the room-temperature electrical noise. Below the critical temperature of $\sim600$~mK, the Ti/Al leads become superconducting, resulting in the formation of three different superconductor-graphene-superconductor (SGS) junctions.

Figure~\ref{fig:set}(b) shows the gate voltage ($V_g$) dependence of the differential resistance ($R_d$) of the graphene sections L-C, C-R and of the whole flake (L-R), at 4~K. The entire section L-R presents a peak in $R_d$ at a gate voltage $V_{\rm CNP}=-11$~V. This resistance peak is associated with a minimum charge carrier density and takes place at the charge neutrality point (CNP). The negative value of $V_{\rm CNP}$ indicates that the flake is {\it n}-doped in the absence of gate voltage. The change in the resistance is smaller in the {\it p}-doped region than in the {\it n}-doped one. This asymmetry comes from the {\it n}-doping by the Ti/Al leads \cite{lee, muellert, Janne2} and the resulting formation of {\it p-n} junctions when bulk of the graphene is {\it p}-doped by the gate for $V_g<V_{\rm CNP}$. For $V_g>V_{\rm CNP}$, we estimate from the Drude model a mobility of $3500$~cm$^2$V$^{-1}$s$^{-1}$ and a mean-free-path of $70$~nm at $V_g=30$~V. The electrical transport through the studied sections is therefore diffusive. The resistance $R_N$ of the short section (C-R) is continuously decreasing with $V_g$ and no resistance peak is observed within the investigated gate voltage range. The short section is thus always {\it n}-doped. This is because the contacts affect the sample on a scale of micrometers so that the average doping in the short section is stronger.

\begin{figure}[t!]
\centering
\includegraphics[width=0.9\columnwidth]{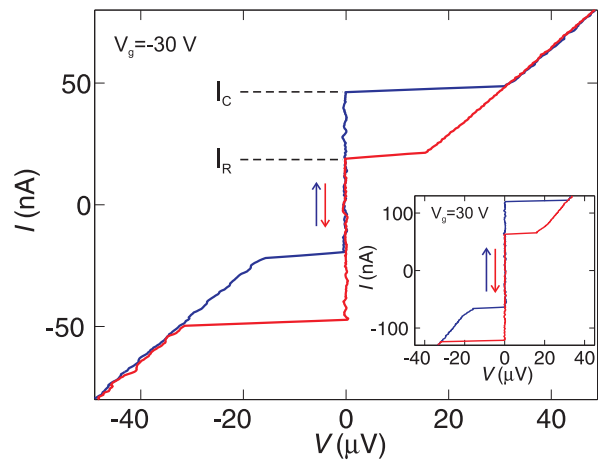}
\caption{(color online). Current-voltage characteristic of the short section C-R at $V_g=-30$~V obtained by successively sweeping up (blue curve) and down (red curve) the bias current at 80~mK. The SGS junction jumps from the zero-voltage state into the resistive state at the critical current $I_c$ and comes back into its initial state at the retrapping current $I_r$. Inset: $IV$ curve of the short section C-R at $V_g=30$~V and T=80~mK.}
\label{fig:IVs_diff_gate}
\end{figure}

We now focus on the current-voltage ($IV$) characteristics of the long and short sections measured at 80~mK for different gate voltages. The $IV$ curves generally contain a supercurrent branch characterized by a zero-voltage state when the bias current is kept below the critical current $I_c$ (see Fig.~\ref{fig:IVs_diff_gate}). At $I_c$, the SGS junction jumps into the resistive state. The initial superconducting state is recovered when the current is biased below the so-called retrapping current $I_r$. For the short sample, at 80~mK, $I_r$ always differ from $I_c$ leading to a hysteretic $IV$ curve. The critical current $I_c$ depends on the gate voltage and decreases when the normal resistance increases as seen in Fig.~\ref{fig:set}. When the gate voltage is tuned from $-30$ to $30$~V, the critical current increases from $47$ to $120$~nA and the normal resistance is lowered from 708 to 337~$\Omega$, respectively. 

%
%
\begin{figure}[t!]
\centering
\includegraphics[width=0.9\columnwidth]{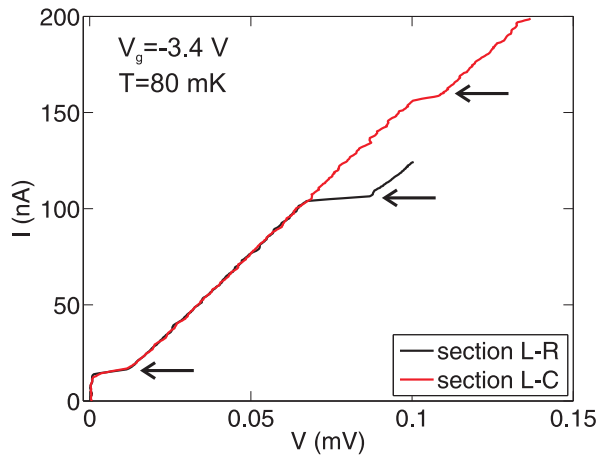}
\caption{(color online). $IV$ curves of sections L-R and L-C at $V_g=-3.4$~V and $T=80$~mK. Three voltage transitions are highlighted by the arrows.}
\label{fig:IV_long_section}
\end{figure}

Fig.~\ref{fig:IV_long_section} shows the current-voltage characteristics of the sections L-C and L-R  at $V_g=-3.4$~V. Both $IV$ curves present a supercurrent branch with the same critical current of 16~nA, and identical differential resistance values (650~$\Omega$) at sufficiently low bias. This is understood by the presence of a supercurrent through the short section (C-R) keeping leads C and R at the same potential. The transition of the short sample into the resistive state is identified by a second voltage jump. This jump takes place at a critical current of $156$~nA for the section L-C and $104$~nA for the section L-R. The differential resistances increase to 685 and 810~$\Omega$ for sections L-C and L-R, respectively.

\begin{figure}[t!]
\centering
\includegraphics[width=0.9\columnwidth]{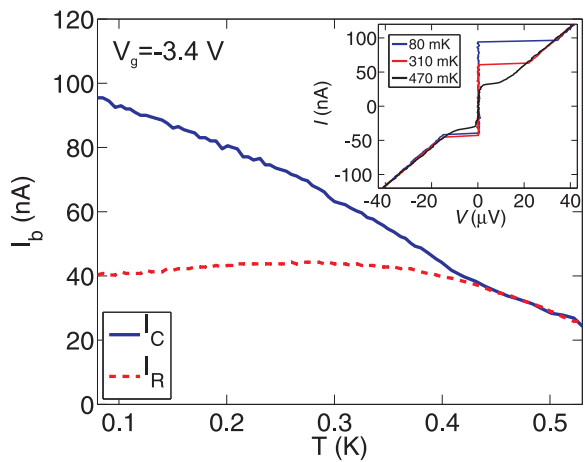}
\caption{(color online). Temperature dependence of the critical (solid line) and retrapping (dashed line) currents at $V_g=-3.4$~V. Inset: Current-voltage characteristic of the short section C-R at $V_g=-3.4$~V for different temperatures. The current is swept up from -120 to 120~nA.}
\label{fig:IVs_diff_temp}
\end{figure}

The $IV$ characteristics of our SGS junctions are strongly changing with temperature. Fig.~\ref{fig:IVs_diff_temp} shows the temperature dependence of the $IV$ curve of the short section (C-R) at $V_g=-3.4$~V. The critical current decreases with increasing the temperature. The retrapping current remains almost constant until 0.3 K and then goes down. The hysteresis of the $IV$ curve is reduced as the temperature goes up and disappears at $0.45$~K. Similar results are found at different gate voltages, and also in the long section L-C. The temperature dependence of the critical current is used in Sec.~\ref{sec:res} to extract the charge carrier mean free path. From the shape of this dependence we can already tell that we are in the long junction regime, where the length of the junction $L$ is longer than the superconducting coherence length $\xi$. Most importantly, the strong temperature dependence of the critical current allows us to use SGS junctions as electronic thermometers.

\begin{figure}[t!]
\centering
\includegraphics[width=0.9\columnwidth]{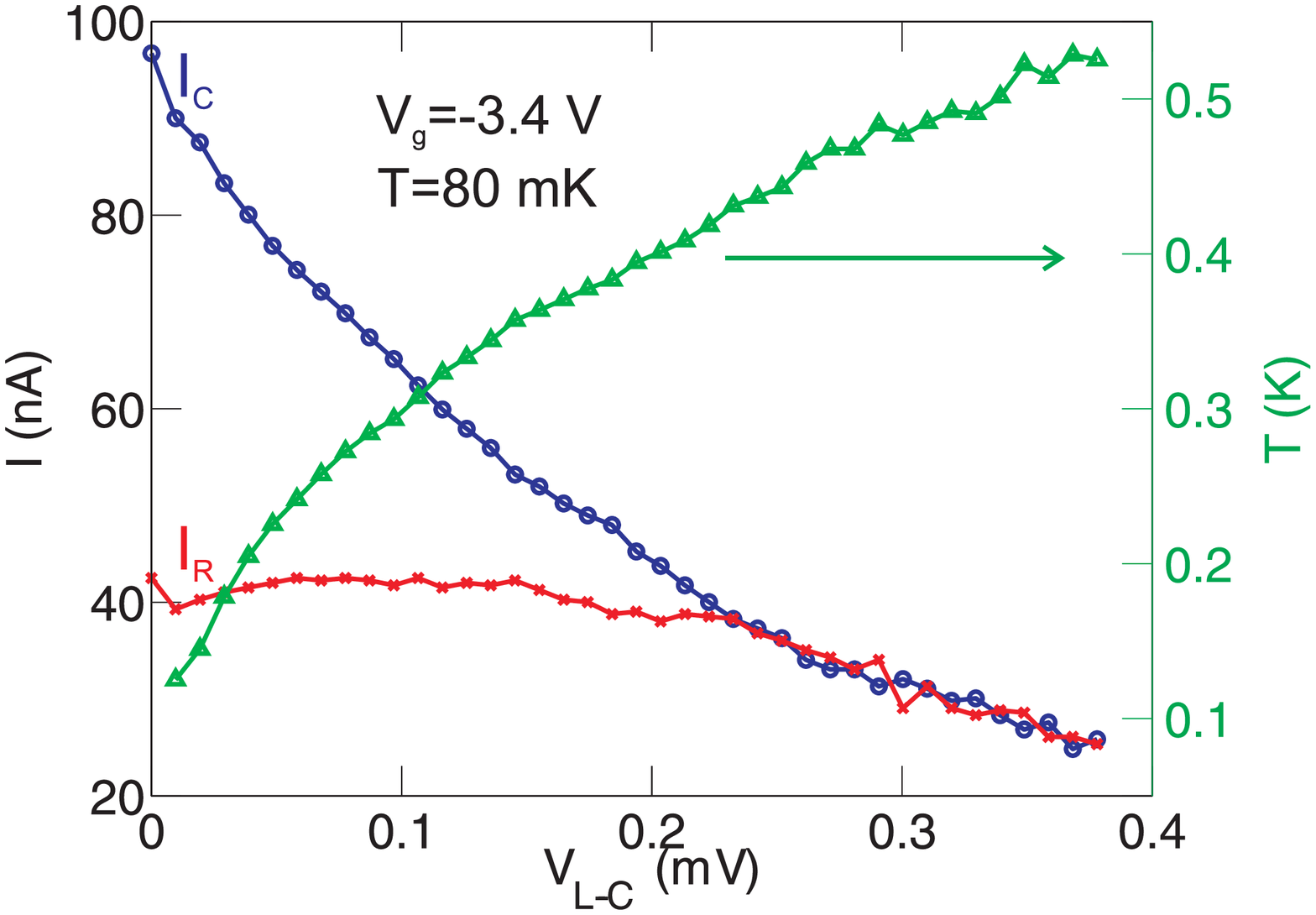}
\caption{(color online). Critical (circle points) and retrapping (cross points) currents as a function of the voltage $V_{\rm L-C}$ across the section L-C. The triangle points corresponds to the deduced electronic temperature.}
\label{fig:Ic_vs_Vlc}
\end{figure}


Keeping the bath temperature at 80~mK, the electronic temperature can be changed by injecting a dissipative current in between leads L and C. As shown in Fig.~\ref{fig:Ic_vs_Vlc} for $V_g=-3.4$~V, the critical current decreases in the short section (C-R) as the voltage $V_{\rm L-C}$ across the long section L-C goes up. The retrapping current remains initially almost constant at low voltages and decreases at around $0.15$~mV. By assuming that the electronic distribution follows a Fermi distribution, we can directly relate the critical current to the electonic temperature. We find that the temperature amounts to around $0.52$~K at $V_{\rm L-C}=0.35$~mV. However, as we show in Sec.~\ref{sec:thry}, the electronic distribution function may differ from a Fermi distribution in which case the electronic temperature is not properly defined. Consequently, the electronic temperature directly deduced from the measurement of $I_c$ is only an effective temperature (see Eq.~\eqref{eq:Teff} below).

\section{Theoretical model}\label{sec:thry}
\begin{figure}[htb]
\centering
\includegraphics[width=0.9\columnwidth]{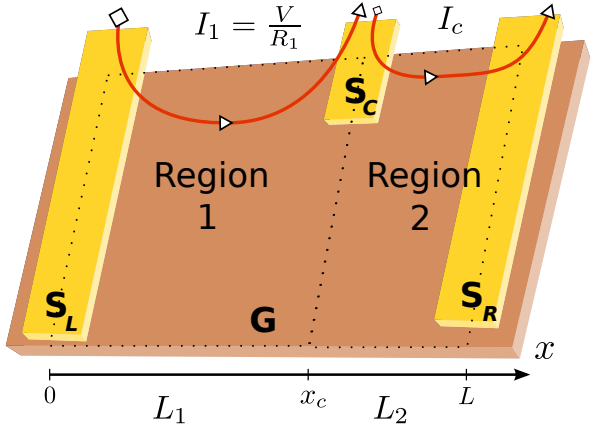}
\caption{(color online). Sketch of the model with graphene flake divided into two regions: the heater (region 1) and thermometer (region 2). The relative sizes of the regions are determined from critical-current measurements as explained in the text.}
\label{fig:graph}
\end{figure}

We compose a model where a graphene flake is divided into two parts by superconducting electrodes with energy gaps $\Delta$ and treat the system as effectively one-dimensional with the essential dimension aligned along the $x$-direction. This setup is illustrated in Fig.~\ref{fig:graph} where the two distinct regions are numbered as 1 and 2.

\subsection{Distribution function and supercurrent}\label{sec:distrib}
Physical observables, such as the supercurrent, can now be determined from the electron distribution function $f(\epsilon)$ which is a function of both energy $\epsilon$ and position $x$, although below the latter is not explicitly written down. The distribution function satisfies the time-independent diffusion equation (disregarding the proximity effect)
\begin{equation}\label{eq:diff}
-D\frac{\partial^2 f(\epsilon)}{\partial x^2}=I(f(\epsilon)).
\end{equation}
Here $D$ is the diffusion constant, related to the Fermi velocity $v_F=10^6$ m/s and the transport relaxation time $\tau$ for elastic scattering by
\begin{equation}\label{eq:D}
 D=v_F^2\tau/2,
\end{equation}
and $I(f)$ is the collision integral for the inelastic processes in the flake (most importantly e-e and e-ph scattering).
At the graphene-superconductor interfaces we have two kinds of boundary conditions depending on whether we are at the end points $x=0$ and $x=L$ or at the boundary of regions 1 and 2, at $x=x_c$. At the end points \cite{PauliTero}
\begin{subequations}\label{eq:bcend}
\begin{equation}\label{eq:eq}
f(\epsilon)=f_{L/R}(\epsilon), \quad |\epsilon|>\Delta
\end{equation}
 \begin{equation}\label{eq:balance}
f(\mu+\epsilon)=1-f(\mu-\epsilon),\quad |\epsilon|<\Delta
 \end{equation}
\begin{equation}\label{eq:heatblock}
\partial_x f(\mu+\epsilon)=\partial_x f(\mu-\epsilon),\quad |\epsilon|<\Delta.
\end{equation}
\end{subequations}
Here, $f_{L/R}(\epsilon)$ is the equilibrium Fermi function at the corresponding potential, i.e.,\ $f_L(\epsilon)=f_0(\epsilon-eV)$ and $f_R(\epsilon)=f_0(\epsilon)$ as $f_0(\epsilon)\equiv \left[\exp{\left[(\epsilon-\mu)/k_BT_{\rm bath}\right]}+1\right]^{-1}$ at base temperature $T_{\rm bath}$ and chemical potential $\mu$ with respect to the Dirac point. Below the gap, these boundary conditions conserve the balance of positive and negative charge excitations \eqref{eq:balance} and ensure that there is no energy current entering the superconductor \eqref{eq:heatblock}, due to Andreev reflection. Above the gap, we simply have continuity of the particle distribution across the contact since the interface is assumed transparent \eqref{eq:eq}. At the dividing superconductor at $x=x_c$, we require
\begin{subequations}\label{eq:bcmid}
\begin{equation}\label{eq:bcmida}
\left.f(\epsilon)\right|_{x=x_{c-}}=\left.f(\epsilon)\right|_{x=x_{c+}}
\end{equation}
\begin{equation}
f(\epsilon)=f_0(\epsilon), \quad |\epsilon|>\Delta
\end{equation}
\begin{equation} 
f(\mu+\epsilon)=1-f(\mu-\epsilon),\quad |\epsilon|<\Delta
\end{equation}
\begin{equation}\label{eq:bccont}
\begin{split}
\partial_x\left.f(\mu+\epsilon)\right|_{x=x_{c-}}-\partial_x\left.f(\mu-\epsilon)\right|_{x=x_{c-}}=&\\
\partial_x\left.f(\mu+\epsilon)\right|_{x=x_{c+}}-\partial_x\left.f(\mu-\epsilon)\right|_{x=x_{c+}},&\quad |\epsilon|<\Delta.
\end{split}
\end{equation}
\end{subequations}
In addition to the requirement of continuity across $x=x_c$, Eq.~\eqref{eq:bcmida}, we have the same condititions as above apart from Eq.~\eqref{eq:bccont} which requires that also the energy current is conserved across $x=x_c$. Physically, we assume that the superconductor on top of the graphene flake acts as an electrode for the high-energy electrons and does not noticeably affect the low-energy ones. We also assume above and in the following that $\Delta$ is constant: independent of position or heating in the system.

The degree of inelastic scattering determines the state of the system which, in the presence of superconductors, can become a quite complicated nonequilibrium state when e-e relaxation is incomplete. \cite{MAR-S} On the other hand, when e-e relaxation is complete, two separate alternatives are possible: The system can either be in the equilibrium state $f_0(\epsilon)$ with bath temperature $T_{\rm bath}$ or in a so-called quasiequilibrium state $f_0(\epsilon)$ with electron temperature $T_e>T_{\rm bath}$ depending on whether the region in question is heated in one way or another. In our configuration, regions 1 and 2 are separated at $x=x_c$ by a grounding superconductor effective at $\epsilon>\Delta$. Therefore, for the ideal case of complete e-e relaxation, region 2 remains in equilibrium at temperature $T_{\rm bath}$ and region 1 in quasiequilibrium with $T_e$ determined by the heating voltage $V$. However, when e-e relaxation is incomplete, the presence of the heat link between the regions makes it a priori possible that the system in region 2 is in any of these states: equilibrium, quasiequilibrium or nonequilibrium. Irrespective of the state of the system, we can illustrate the heat distribution in the system by defining a (local) effective electron temperature
\begin{equation}\label{eq:Teff}
k_BT_e=\frac{\sqrt{6}}{\pi}\sqrt{\int_{-\infty}^\infty\epsilon\left[f(\epsilon)-1+\theta(\epsilon)\right]d\epsilon}
\end{equation}
where $\theta(\epsilon)$ is the Heaviside step function. This equates the energy in the system to the thermal energy.

For the quantitative results, we solve the distribution function $f(\epsilon)$ from the diffusion equation Eq.~\eqref{eq:diff} and when $V=0$, we use the equilibrium Fermi function $f(\epsilon)=f_0(\epsilon)$ with a given bath temperature $T_{\rm bath}$. In both cases, the supercurrent through graphene is \cite{Belzig}
\begin{equation}\label{eq:IS}
I_S(\phi)=\frac{C}{eR_N}\int_0^\infty d\epsilon j_S(\epsilon,\phi)[1-2f(\epsilon)],
\end{equation}
where we average $f(\epsilon)$ over the $x$-coordinate. Here, $C\lesssim1$ is a prefactor of the order unity describing the imperfections in the measurement. The spectral supercurrent $j_S(\epsilon,\phi)$ of a diffusive SNS system is defined in Ref.~\onlinecite{Belzig} and we determine it numerically at the phase $\phi=0.6\pi$, which gives a fair approximation for the low-temperature critical current $I_c\equiv \max I_S(\phi)\approx I_S(0.6\pi)$.

We note that for graphene the diffusion constant $D$ can depend strongly on the type of disorder and on doping.  For screened Coulomb impurities,\cite{adam,Janne2} $D\propto\tau\propto |\mu|$.  Below, $D$ is essentially fit to the experiments, so that we do not need to specify the nature of the scatterers. The use of a semiclassical diffusion approach has some limitations for the description of graphene, however. The experimentally found finite minimal conductivity at the CNP cannot be explained without quantum-mechanical effects and a more detailed description of the impurities and the nonuniform doping effects they may introduce (charge puddles). These are not an issue if the graphene is strongly {\it n} or {\it p} doped. In our case, for $V_g>V_{\rm CNP}$ the doping is of {\it n} type everywhere, but as mentioned above, for $V_g<V_{\rm CNP}$ {\it p-n} junctions are expected to emerge close to the contacts.\cite{lee,muellert,Janne2} For these reasons, and because of the assumed one-dimensionality, our description should be viewed only as an effective model, in particular when $V_g<V_{\rm CNP}$. 

\subsection{Inelastic interactions: electron-phonon}\label{sec:e-ph}
We start by estimating the strength of e-ph contribution in the inelastic collision integral. For this, we deem it sufficient to pay attention only to acoustic phonons for the range of temperatures and voltages relevant to our experiments ($T,eV/k_B\sim 1$~K $\cong 0.1$~meV). The collision integral for acoustic phonons in graphene has been derived in Ref.~\onlinecite{Janne} but here we only need the e-ph power discussed in Refs.~\onlinecite{Janne, giazotto06, kubakaddi09, bistritzer09}. We assume the distribution $f(\epsilon)$ is sufficiently well defined by the effective electron temperature $T_e$ of Eq.~\eqref{eq:Teff}. Then, in the limit $(c/v_F)|\mu|\gg k_BT_e$, where $c$ is the sound velocity, the e-ph power is 
\begin{equation}\label{eq:Peph}
P_{\rm e-ph}=\Sigma(\mu) A(T_e^4-T_{\rm bath}^4),
\end{equation}
with a $\mu$-dependent interaction constant $\Sigma(\mu)$ and the area of the flake $A$. This power law is applicable up to very close to the Dirac point since the condition $\mu=\hbar v_Fk_F=\hbar v_F\sqrt{\pi\epsilon_r\epsilon_0\delta V_g/ed}\gg k_BT_ev_F/c$ translates to
\begin{equation}
\delta V_g\gg\left(\frac{k_BT_e}{\hbar c}\right)^2\frac{ed}{\pi\epsilon_r\epsilon_0}
\end{equation}
for the distance from the Dirac point in terms of gate voltage $\delta V_g\equiv |V_g-V_{\rm CNP}|$. For $T_e\sim 1$~K, thickness of the gate oxide $d=250$~nm, relative permittivity $\epsilon_r=4$ and $c=0.02v_F=2\times 10^4$~m/s, this becomes $\delta V_g\gg 15$~mV. In our experimental data the minimal $\delta V_g$ is about $1.5$ V, and so the condition is not violated. This estimate neglects the effect of charge puddles close to the CNP, but we expect Eq.~\eqref{eq:Peph} to hold whenever the use of our semiclassical approach is justified.

The total power injected into the system is $P_{\rm in}=V^2/R_N$ and if electron-phonon coupling is absent, all of this escapes into the leads. If, furthermore, e-e interactions are neglected, the electron distribution function is comprised of discrete steps due to multiple Andreev reflections \cite{MAR-S} so that highest local effective temperature of the system is roughly $k_BT_e^{\rm max}\approx(\Delta+eV)/4$. \cite{myfoot} As a result, energy escape takes place also at $eV\ll\Delta$. In the presence of e-e interactions, $T_e$ is reduced from this value so that for strong interactions, $k_BT_e^{\rm max}\approx eV/4$ as in a diffusive conductor without superconductivity (Notice that, at high voltages, a hot spot is formed in the middle of the normal-conducting region 1, so that $T_e^{\rm max}$ becomes high while the average $T_e$ remains much lower). For $T_e\gg T_{\rm bath}$
\begin{equation}
 \frac{P_{\rm e-ph}}{P_{\rm in}}=\frac{\Sigma AT_e^4R_N}{V^2},
\end{equation}
and we see that the relative importance of e-ph interaction increases in two cases: first with increasing $eV$ and second with $eV$ decreasing below $\Delta$ when e-e interaction is absent. We evaluate the both possibilities to obtain a range for $eV$ where $P_{\rm e-ph}$ is significant.

The value for the e-ph interaction constant has not been measured but we obtain a theoretical upper estimate $\Sigma=k_B^4\mathcal{D}^2\pi^2|\mu|/(60\hbar^5\rho v_F^3c^3)<6.9\times 10^{-3}$~W/K$^4$m$^2$ for graphene mass density $\rho=0.76$~mg/m$^2$, deformation potential constant $\mathcal{D}=10$~eV (see, for example, Ref.~\onlinecite{borysenko}) and chemical potential $\mu=0.22$~eV, corresponding to $V_g=+30$~V. This is roughly an order of magnitude smaller than e-ph interaction in metals, \cite{giazotto06} when the reduced dimensionality of graphene is taken into account using thickness $\sim 1$~\AA. Setting $A\approx 2.8$~$\mu$m $\times 4.0$~$\mu$m, $\Delta/k_B\approx 1.12$~K, and $R_N\approx 1$~k$\Omega$, yields $\frac{P_{\rm e-ph}}{P_{\rm in}} < 0.013\times (k_BT_e/\Delta)^4/(eV/\Delta)^2$. We then have the result:
\begin{equation}
\frac{P_{\rm e-ph}}{P_{\rm in}}<5\times 10^{-5} \left(\frac{eV}{\Delta}\right)^2\left(1+\frac{\Delta}{eV}\right)^4,
\end{equation}
assuming $k_BT_e=(\Delta+eV)/4$. The ratio is above 1\% when $eV<0.083\Delta$ or $eV>12\Delta$, while our measurements are focused on the range $eV=0.5\ldots 5\Delta$. We emphasize that in this estimate we used the hot-spot temperature which is much larger than the average $T_e$ at high voltages. On the other hand, at low voltages, we assumed a total absence of e-e interaction resulting in a high $T_e$ of the order $\Delta$. Generally, $T_e$ can be expected to be even smaller than estimated above and we conclude that for our experiment e-ph coupling can be neglected.

\subsection{Inelastic interactions: electron-electron}\label{sec:e-e}
Coulomb interactions in graphene have been discussed, for example, in Refs.~\onlinecite{mirlin, mueller, tse2}.  However, most of the existing results are for clean, charge-neutral graphene, where  the golden-rule collision integrals are furthermore plaqued by divergences.\cite{mueller}  A full theory of e-e interactions for diffusive graphene that would be valid at both the Dirac point and at finite doping is currently lacking.  In particular it is not known if a well-defined quasiequilibrium state ever exists in diffusive graphene biased far from equilibrium, although it is often a convenient assumption.\cite{Janne2} Since the interband relaxation due to e-e collisions is expected to be weak,\cite{mirlin,foster09,balev} electrons and holes (or electrons in the conduction and valence bands) may in any case have to be treated separately.\cite{balev}

As explained above, our semiclassical approach restricts our calculation in principle to the strongly-doped regime, where only 
one charge carrier is dominant. In this case the system may be expected to behave somewhat similarly to other disordered 
two-dimensional conductors. Thus, a reasonable starting point for an effective description is the Altshuler-Aronov theory\cite{AA} 
for diffusive normal metals. The collision integral for a well-screened diffusive wire in two dimensions is \cite{AA, RamSmth}
\begin{equation}\label{eq:Iee}
\begin{split}
I_{e-e}(f(\epsilon))=
\kappa_{e-e}&\int_{-\infty}^\infty d(\hbar\omega)\;
\int_{-\infty}^\infty d\epsilon'(\hbar\omega)^{-1}\\
&\times\left[I^{\rm in}(\omega,\epsilon,\epsilon')-I^{\rm out}(\omega,\epsilon,\epsilon')\right]
\end{split}
\end{equation}
with 
$I^{\rm in}(\omega,\epsilon,\epsilon')=[1-f(\epsilon)][1-f(\epsilon')]f(\epsilon-\hbar\omega)f(\epsilon'+\hbar\omega)$
and $I^{\rm out}(\omega,\epsilon,\epsilon')=f(\epsilon)f(\epsilon')[1-f(\epsilon-\hbar\omega)][1-f(\epsilon'+\hbar\omega)]$. 
The prefactor is given by $\kappa_{e-e}=1/(8|\mu|\tau)$ and while the distribution functions are assumed position-dependent, we disregard any such dependence in $\kappa_{e-e}$ for simplicity.

%
%
We can also estimate the e-e relaxation rate, which reads \cite{RamSmth2}
\begin{equation}\label{eq:tee}
\frac{1}{\tau_{e-e}}=\frac{k_BT_e}{2m_eD}\ln\frac{T_1}{T_e}
\end{equation}
in energy-averaged form valid at low-energies $\sim k_BT_e$. The result is obtained from the collision integral, where an energy cut-off is required to remove a logarithmic divergence at small $\omega$, characteristic to two-dimensional systems.\cite{abraham} The cut-off energy is $\hbar\omega_0=4k_BT_e^3/T_1^2$ and this gives $T_1=m_e^2D^3e^4N_F^2/\hbar^2k_B\varepsilon_0^2$, also for Eq.~\eqref{eq:tee}. Here, $m_e$ is the electron mass, $\varepsilon_0$ the vacuum permittivity and $N_F$ the density of states at the Fermi level. For graphene, the latter becomes \cite{Grev}
\begin{equation}\label{eq:NF}
N_F=\frac{2}{\pi}\frac{|\mu|}{(\hbar v_F)^2},
\end{equation}
including spin degeneracy.

\subsection{Final model in dimensionless form}
With the e-e interaction as the sole contributor to the inelastic relaxation, we now cast the diffusion equation Eq.~\eqref{eq:diff} in a dimensionless form directly applicable for numerical solution:
\begin{equation}\label{eq:diffnodim}
\partial_{\tilde x}^2f=-K_{e-e}\tilde I_{e-e},
\end{equation}
where $\tilde x=x/L$ with $L$ denoting the total length of the graphene flake. The dimensionless parameter $K_{e-e}$ describing the strength of e-e interaction in two dimensions is derived from $\kappa_{e-e}$ given above and expressed in terms of experimentally relevant parameters. When all energies are normalized by $\Delta$,
\begin{equation}\label{eq:Kee}
 K_{e-e}=\frac{L^2}{D}\Delta\kappa_{e-e}=\frac 14\frac{R_N}{R_Q}\frac{\Delta}{E_{\rm Th}}\frac{w}{L},
\end{equation}
where we have used the formula $\sigma=e^2N_FD=L/R_Nw$ for conductivity and Eqs.~\eqref{eq:D} and \eqref{eq:NF} for $D$ and $N_F$.
Here, $R_N$ is the normal-state resistance of the graphene flake, $R_Q\equiv h/e^2$ the quantum of resistance, $E_{\rm Th}=\hbar D/L^2$ the Thouless energy, $w$ the width of the flake and $L$ its length. Note that $K_{e-e}$ now depends on the dimensions of the sample and we define it, together with the other extensive materials parameters, here for the whole flake. Since $E_{\rm Th}\propto D$ and $R_N\propto D^{-1}$, the parameter $K_{e-e}$ is predicted to scale as $K_{e-e}\propto D^{-2}\propto R_N^2$.

We note again that while $K_{e-e}$ does not depend explicitly on $\mu$, such dependence is in principle present through the diffusion constant $D(\mu)$ (and hence $R_N$ and $E_{\rm Th}$). Below, we use only the single parameter $K_{e-e}$ to characterize the e-e interaction and extract it from the experiments.

\section{Results}\label{sec:res}
\subsection{Electron-electron strength}
In order to access the e-e scattering strength $K_{e-e}$, we make two measurements. In the first one, the bath temperature is varied. In the second, an injection voltage at the left superconductor is used to heat up the graphene flake. The critical current and the retrapping current in region 2 are then measured as functions of temperature and voltage, respectively. In the first measurement, the system is in thermal equilibrium, whereas in the second one, it can be in a nonequilibrium state. Therefore, we use the first measurement to determine $E_{\rm Th}$ in regions 1 and 2, and the second measurement to compare the experiments with our theoretical model.

\begin{table}
\caption{Graphene properties determined for different gate voltages. The results for $E_{\rm Th}$ and $C$ (Eq.~\eqref{eq:IS}) are based on critical-current measurements for which we have data for all gate voltages apart from $V_g=-30$~V and $V_g=-3.4$~V in region 1.}
\label{table:ETh}
\begin{center}
    \begin{tabular}{ | c | c || c | c | c || c | c | c |}
    \hline
    & & \multicolumn{3}{|l||}{Region 1} & \multicolumn{3}{| l|}{Region 2} \\
    $V_g$ & $\mu$(meV) & $R_N$ & $E_{\rm Th}/\Delta$ & $C$ & $R_N$ & $E_{\rm Th}/\Delta$ & $C$ \\ \hline
    $-30~V$ & 150 & $760~\Omega$ & N/A & N/A & $708~\Omega$ & $1/3.5$ & 0.33 \\
    $-9.5~V$ & 42 & $1100~\Omega$ & $1/6.4$ & 0.11 & $547~\Omega$ & $1/2.35$ & 0.42 \\
    $-3.4~V$ & 96 & $650~\Omega$ & N/A & N/A & $467~\Omega$ & $1/2.5$ & 0.39 \\
    $+30~V$ & 220 & $330~\Omega$ & $1/7.7$ & 0.18 & $337~\Omega$ & $1/2.7$ & 0.36 \\
    \hline
    \end{tabular}
\end{center}
\end{table}
\begin{figure}[t!]
\centering
\includegraphics[width=0.9\columnwidth]{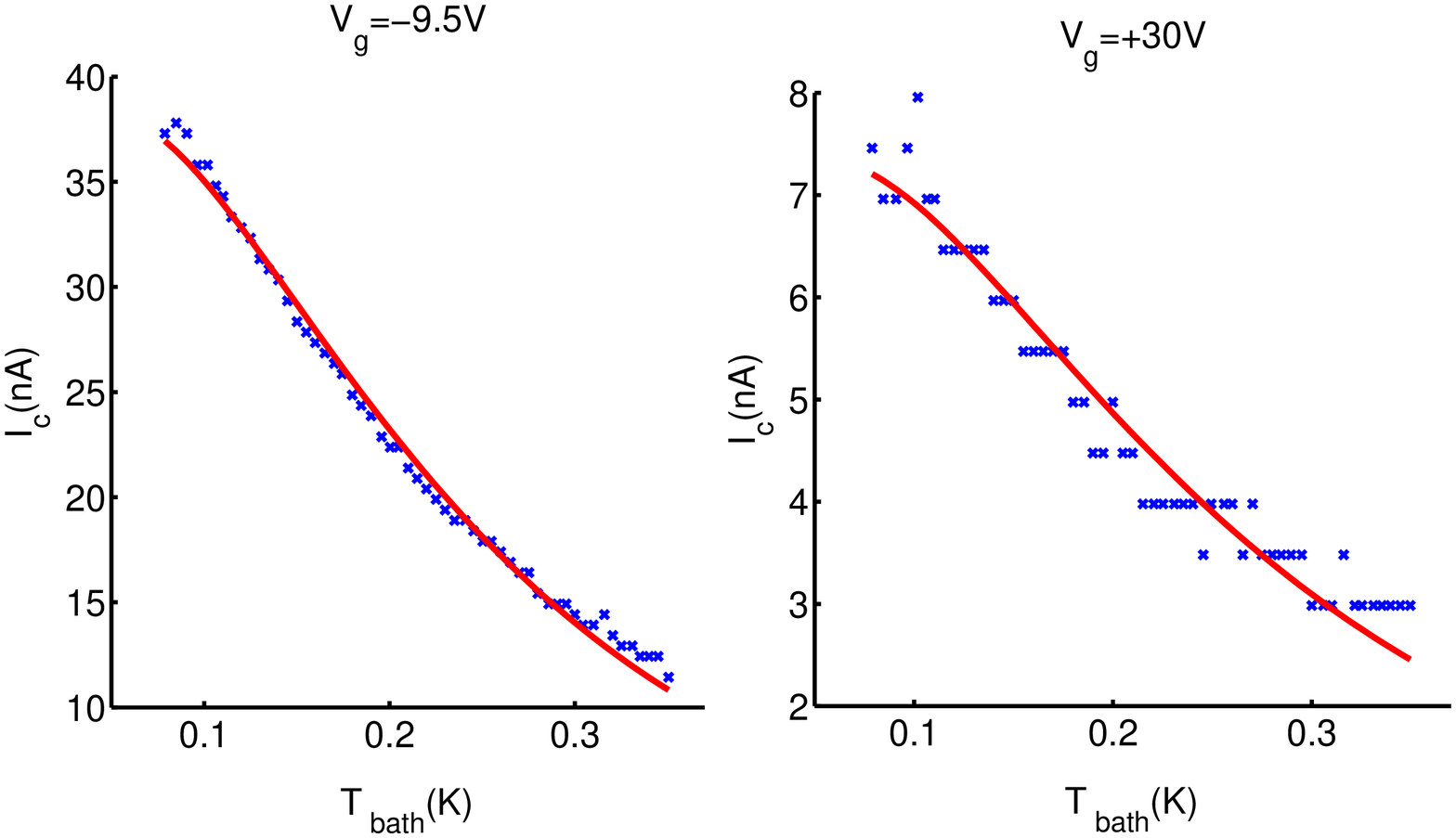}
\caption{(color online). Critical current in region 1 as a function of bath temperature. Solid lines are the fits to the theory, using $E_{\rm Th}$ and $C$ as the fitting parameters.}
\label{fig:EThfit_long}
\end{figure}
\begin{figure}[t!]
\centering
\includegraphics[width=0.9\columnwidth]{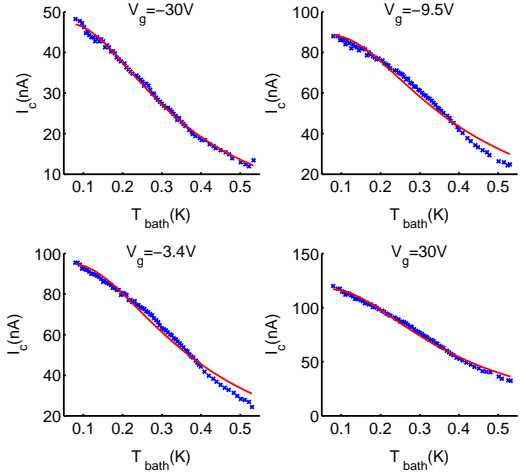}
\caption{(color online). Critical current in region 2 as a function of bath temperature. Solid lines are the fits to the theory, using $E_{\rm Th}$ and $C$ as the fitting parameters.}
\label{fig:EThfit_short}
\end{figure}
In thermal equilibrium, we calculate the theoretical value for $I_c$ simply by using an equilibrium function $f_0$ at temperature $T_{\rm bath}$ in Eq~\eqref{eq:IS}. Here and below, $\Delta=100\;\mu$eV $\cong 1.1$~K so that the critical temperature $T_c\sim$ 0.6~K. Since $j_S(\epsilon,\phi)$ in Eq.~\eqref{eq:IS} is dependent on $E_{\rm Th}$, we may fit the calculated value of $I_c$ to the measured one at different $V_g$ (Figs.~\ref{fig:EThfit_long} and~\ref{fig:EThfit_short}), to obtain the values in Table~\ref{table:ETh}. These values for $E_{\rm Th}$ are subsequently used to determine the effective lengths $L_1$ and $L_2$ of regions 1 and 2, respectively. We assume that the diffusion constant $D$ is constant throughout the graphene flake so that
\begin{equation}
\frac{L_1}{L_2}=\sqrt{\frac{E_{\rm Th}({\rm region\;2})}{E_{\rm Th}({\rm region\;1)}}}\approx \frac 53.
\end{equation}
This approximate value is used in the simulation and assuming $L=L_1+L_2$, we have $L_1=1.75$~$\mu$m and $L_2=1.05$~$\mu$m. For consistency, we may also check the elastic scattering length $l=2L^2E_{\rm Th}/\hbar v_F$ resulting from the values we obtained for $E_{\rm Th}$ and depending on $V_g$, we have $l=90\ldots 140$~nm. At $V_g=30$~V, the value $l=120$~nm is relatively close to the experimentally determined $l=70$~nm and the difference can be due to inaccuracy in determining the effective length.

\begin{table}
\caption{Parameter values determined for the graphene flake. Resistance is calculated as the sum of $R_N$ in region 1 and 2, whereas Thouless energy is obtained assuming that the diffusion constant determined from the $I_c$-measurements in region 2 is homogeneous throughout the whole flake. The theoretical value for $K_{e-e}$ is calculated from Eq.~\eqref{eq:Kee}.}
\label{table:Kee}
\begin{center}
    \begin{tabular}{ | c || c | c | c | c | c |}
    \hline
    \multicolumn{6}{|l|}{The whole flake}\\
    $V_g$ & $R_N$ & $E_{\rm Th}/\Delta$ & $D$ & $K_{e-e}^{\rm fit}$ & $K_{e-e}^{\rm theory}$\\ \hline
    $-30~V$ & $1468~\Omega$ & $1/24.9$ & 0.046~m$^2$/s & 30.8 & 0.51 \\
    $-9.5~V$ & $1647~\Omega$ & $1/16.6$ & 0.069~m$^2$/s & 51.9 & 0.38 \\
    $-3.4~V$ & $1117~\Omega$ & $1/17.8$ & 0.065~m$^2$/s & 26.6 & 0.28 \\
    $+30~V$ & $667~\Omega$ & $1/19.5$ & 0.060~m$^2$/s & 7.6 & 0.18 \\
    \hline
    \end{tabular}
\end{center}
\end{table}
\begin{figure}[t!]
\centering
\includegraphics[width=0.9\columnwidth]{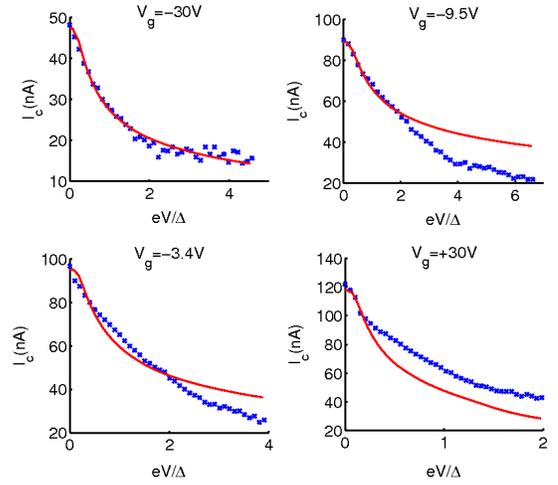}
\caption{(color online). Critical current in region 2 as a function of heater voltage. Solid lines are the fits to the theory, using $K_{e-e}$ as the fitting parameter at a single point $eV<\Delta$ (The exact value depends on the gate voltage).}
\label{fig:IcV}
\end{figure}
In the second experiment we use the heater voltage $V$ at $T_{\rm bath}=80$~mK and, as a result, $I_c$ in region 2 is dependent on the strength of the e-e interaction $K_{e-e}$. We proceed by fitting the theoretical and experimental results at a single small voltage $eV<\Delta$ with $K_{e-e}$ as the fit parameter. We then use the fitted $K_{e-e}$ to compute the whole $I_c(V)$-curve. The results for different $V_g$ are given in Fig.~\ref{fig:IcV}. The corresponding $K_{e-e}$ are listed in Table~\ref{table:Kee}.

\begin{figure}[t!]
\centering
\includegraphics[width=0.9\columnwidth]{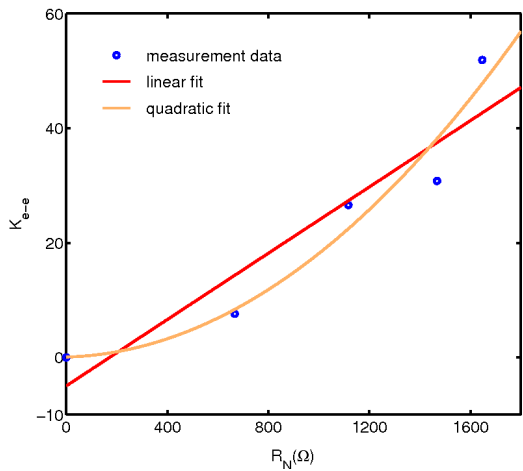}
\caption{(color online). Values for the e-e interaction strength $K_{e-e}$ determined from the experimental data and shown with respect to the normal state resistance of the graphene flake $R_N$.}
\label{fig:KvsR}
\end{figure}
From Eq.~\eqref{eq:Kee} we expect the e-e strength to scale as $K_{e-e}\sim R_N^2$. The values of $K_{e-e}$ extracted from the experiments are shown in Fig.~\ref{fig:KvsR} as a function of $R_N$ from Table~\ref{table:Kee}, assuming additionally that $R_N\rightarrow 0$, $K_{e-e}\rightarrow 0$. The data supports the quadratic behavior.

\subsection{State of the system: Quasi- or nonequilibrium?}
\begin{figure}[t!]
\centering
\includegraphics[width=0.95\columnwidth]{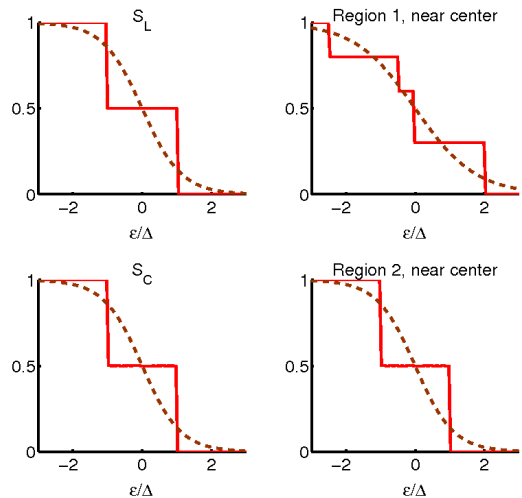}
\caption{(color online). Electron distribution functions $f(\epsilon)$ for a representative case ($V_g=-30$~V, $eV=2.5\Delta$) at different positions of the graphene flake. Here, e-e interactions are neglected, i.e.,\ $K_{e-e}=0$. Dashed line is the quasiequilibrium function $f_0(\epsilon)$ at the effective temperature $T_e$ corresponding to the nonequilibrium function $f(\epsilon)$.}
\label{fig:f0}
\end{figure}

\begin{figure}[t!]
\centering
\includegraphics[width=0.95\columnwidth]{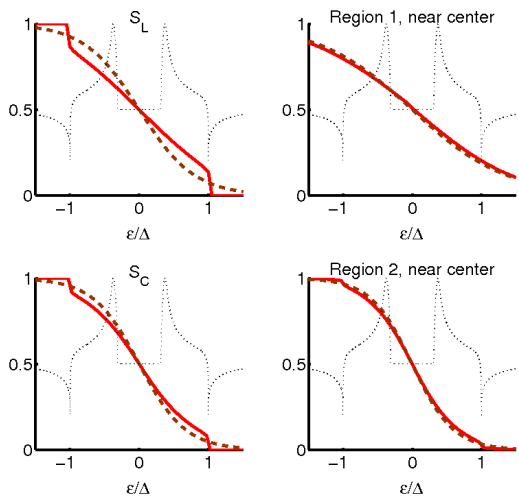}
\caption{(color online). Electron distribution functions $f(\epsilon)$ for a representative case ($V_g=-30$~V, $eV=2.5\Delta$) at different positions of the graphene flake. Here, the strength of the e-e interaction corresponds to the value of Table~\ref{table:Kee} which we obtained by fitting to the experimental data: $K_{e-e}=30.8$. Dashed line is the quasiequilibrium function $f_0(\epsilon)$ at the effective temperature $T_e$ corresponding to the nonequilibrium function $f(\epsilon)$. The spectral supercurrent $j_S(\epsilon)$ (Eq.~\eqref{eq:IS}) is shown as a black dotted line with arbitrary units and centered so that $j_S=0$ when $|\epsilon|<E_{\rm Th}=\Delta/3.5$.}
\label{fig:f1}
\end{figure}

We determine the state of the system and assess how sensitive it is to any changes in $K_{e-e}$ by looking at the electron distribution functions $f(\epsilon)$. In our theoretical model based on metallic e-e interaction, gate voltage has no explicit effect on the results and any gate dependence is due only to such dependence in the materials parameters such as $R_N$. We therefore use the parameter values from the case where $V_g=-30$~V as a representative example. In addition, we set $eV=2.5\Delta$ so that any pecularities due to the nonequilibrium state should be visible both in region 1 and 2. A true nonequilibrium distribution with $K_{e-e}=0$ is given in Fig.~\ref{fig:f0} and we notice that, in the absence of relaxation, $f(\epsilon)$ remains constant in region 2. This happens because the electrostatic potential at the middle superconductor is fixed. The numerically determined $f(\epsilon)$ for $K_{e-e}=30.8$ (Table~\ref{table:Kee}) is shown in Fig.~\ref{fig:f1}. We see that while the electrons clearly have a nonequilibrium distribution near the superconductors, where $f(\epsilon)$ in any case is strongly affected by the boundary conditions of Eqs.~\eqref{eq:bcend} and~\eqref{eq:bcmid}, in the middle of both regions 1 and 2 $f(\epsilon)$ is smoothed very close to a thermal distribution.

\begin{figure}[t!]
\centering
\includegraphics[width=0.9\columnwidth]{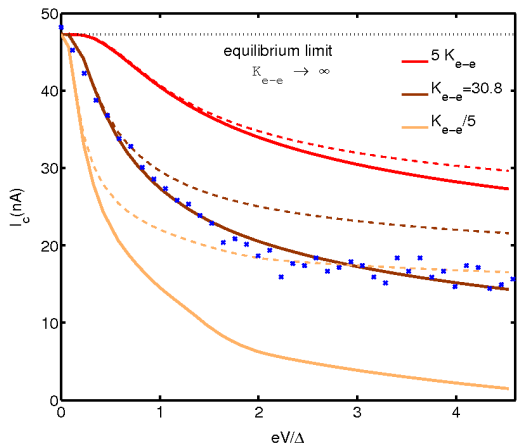}
\caption{(color online). Calculated critical current as a function of voltage for different values of the e-e strength $K_{e-e}$. The solid lines are the results from the nonequilibrium distributions $f(\epsilon)$, whereas the dashed lines demonstrate how $I_c$ differs if quasiequilibrium functions with effective electron temperatures are used instead. Blue crosses correspond to the measured values.}
\label{fig:KtestIc}
\end{figure}

The question now regarding the state of the system is: is $f(\epsilon)$ in region 2 effectively a thermal (quasi)equilibrium distribution? If so, the e-e interaction is weak enough to let some of the energy injected into region 1 leak into region 2, but still so strong that it forces the system in region 2 into a thermal state. To answer this, we calculate $I_c$ as a function of heater voltage for several values of $K_{e-e}$ in Fig.~\ref{fig:KtestIc}. First, the expected value of supercurrent is clearly dependent on the value of $K_{e-e}$ in this range so the measurement can be used to determine $K_{e-e}$ with satisfactory precision. Second, as there is no plateau of constant $I_c$ at small voltages, the system is only in complete equilibrium state at $V=0$. Third, at voltages $eV\gtrsim\Delta$ there is a clear difference between the critical current obtained using actual nonequilibrium distributions and their quasiequilibrium counterparts, i.e.,\ equilibrium functions $f_0(\epsilon)$ with $T_e$ determined from $f(\epsilon)$. This implies that not only electron heating but also the formation of nonequilibrium state affects the observable supercurrent at $eV\gtrsim\Delta$. For voltages smaller than this, a quasiequilibrium description for the system is apt. 

\subsection{Relaxation time}
We finally estimate the e-e relaxation time in our sample. From Eq.~\eqref{eq:tee}, we find $\tau_{e-e}=2\ldots4$~ns at $T_e=T_{\rm bath}=80$~mK and $\tau_{e-e}=400\dots 700$~ps at $T_e=0.5$~K, depending on the gate voltage.
The relaxation rate is expected to be proportional to the electron-electron scatterin strength $K_{e-e}$. Our measurements therefore suggest e-e relaxation times which are smaller than the values above by a factor $K_{\rm e-e}^{\rm fit}/K_{\rm e-e}^{\rm theory}$. Then, at $T_e=T_{\rm bath}$, $\tau_{e-e}=30\ldots 70$~ps and $T_e=0.5$~K, $\tau_{e-e}=5\dots 13$~ps. We are not aware of any comparable results in the low-energy regime. The e-e scattering times have been estimated for graphene in Ref.~\onlinecite{tse2}, but there the graphene is ballistic, the energy scale much larger and the obtained scattering times, consequently, (even) much smaller.

\section{Discussion}\label{sec:dis}
We have measured the strength of e-e interaction in graphene at four different gate voltages. From Eq.~\eqref{eq:Kee} we find the expected values for the parameter $K_{e-e}$ describing the interaction strength as given in Table~\ref{table:Kee}. The measured values are roughly 40-140 times larger than expected from the Altshuler-Aronov theory with the largest differences closer to the Dirac point. Discrepancies between theory and experiments are reported also in metallic wires (Ag), \cite{Huard} but there the differences are sample-specific and only up to a factor of 20, with the measured value larger there as well. Even though the strength of the e-e interaction is stronger than expected, the system is still not thermalized and the incomplete e-e relaxation can be seen at heater voltages $V$ well below the superconducting gap by measuring the critical current. The result at low voltages is seen as an energy leak from the heater junction and increasing temperature in the thermometer region. For $eV>\Delta$, also the electrons in the thermometer are driven into a nonequilibrium state where a thermal description with an effective temperature $T_e$ is no longer enough. In addition to finding the magnitude of the e-e scattering strength, we find that the scattering strength exhibits a significant gate dependence, presumably due to changes in charge density as the gate voltage is varied.

Finally, we note that we have also estimated the interaction strength between electrons and acoustic phonos in our setup with an aim to measure it. However, since the e-e interaction is relatively strong, we expect that $T_e$ is at most of the order of the heater voltage except when $eV\ll\Delta$. Consequently, the expected electron-phonon power at $eV\lesssim\Delta$ becomes even lower than predicted by our estimate in Sec.~\ref{sec:e-ph}. At the other end of the scale, $eV>\Delta$, the effective temperature needs to be increased yet further for a high ratio $P_{\rm e-ph}/P_{\rm in}\sim T_e^4/V^2$. This results in notable heating also in the thermometer region, making a critical current measurement such as the one used here very difficult, unless a strong thermal isolation is established between the heater and the thermometer regions. 

\begin{acknowledgments}
The authors thank Pauli Virtanen and Matti Laakso for helpful discussions. JV gives thanks to Finnish Foundation for Technology Promotion for their support. JKV and TTH acknowledge the support from the Academy of Finland and TTH further acknowledges the funding European Research Council (Grant No. 240362-Heattronics)
\end{acknowledgments}


\begin{thebibliography}{99}
\bibitem{myfoot} Generally, thermal broadening of a distribution function around the Fermi energy has an energy range $\epsilon=-4k_BT\ldots 4k_BT$ giving a factor of 4 in the relation.
\end{thebibliography}
\end{document}